\documentclass[floatfix,showpacs,twocolumn,pre]{revtex4}
\usepackage{epsfig}
\usepackage[dvips]{color}
\usepackage{natbib}

\begin{document}

\title{Models and average properties of scale-free directed networks}

\author{Sebastian Bernhardsson and Petter Minnhagen}\email{minnhagen@tp.umu.se}
\affiliation{Department of Theoretical Physics, Ume{\aa} University,
901 87 Ume{\aa}, Sweden}
\affiliation{Center for Models of Life, Blegdamsvej 17, 2100 Copenhagen, Denmark}

\date{\today}

\begin{abstract}
We extend the merging model for undirected networks by Kim et al.\ [Eur. Phys. J. B 43, 369 (2004)] to directed networks and investigate the emerging scale-free networks. Two versions of the directed merging model, {\em friendly} and {\em hostile merging}, give rise to two distinct network types. We uncover that some non-trivial features of these two network types resemble two levels of a certain randomization/non-specificity in the link reshuffling during network evolution. Furthermore the same features show up, respectively, in metabolic networks and transcriptional networks. We introduce measures that single out the distinguishing features between the two prototype networks, as well as point out features which are beyond the prototypes.
\end{abstract}

\maketitle

\section{Introduction}
The subject of complex networks has recently caused a rapid surge of interest and already quite a few reviews have
surfaced \cite{Albert02,Newman03, Dorog02,Strogatz01,Bocca06}. While a large effort has been devoted to undirected networks, comparatively less attention has so far been payed on modeling directed networks. On the other hand, there exist also many different types of real directed networks from various realm of science\cite{Albert02,Newman03,Dorog02,Strogatz01,Bocca06}. The directed networks-models discussed in the present paper structurally resembles two examples from biology i.e.\ metabolic networks \cite{Jeong00,Barhome,Ma03a,Ma03b,Tanaka05} and transcriptional networks \cite{Hodges99,Hodges98,Yook04}. These two types, while being structurally rather different, both display a broad out-degree distribution.

The scale-freeness of the degree-distributions has been a central issue in the network research \cite{Albert02,Newman03, Dorog02,Strogatz01,Bocca06}. Typical questions we can ask are: Does the feature of broad degree distributions observed in real networks suggest a common cause? Or does the broad distributions arise in a variety of ways, implying that no common cause exists? Can directed networks with scale-free degree distributions be subdivided into "universality classes" based on relations between in and out-degrees in the networks?

In the first step we try to find local organizational update rules, which automatically generate directed networks with broad scale-free-like degree-distributions. We present two such local rules which we term {\em friendly} and {\em hostile} merging. We further characterize the two emerging types of directed scale-free networks by measures connected to the relative amount of and the relations between in- and outgoing links. We notice that directed networks have more characteristic features than undirected ones which make it easier to detect differences between them. Next we introduce two minimalistic random network models which display features reminiscent of the two types of networks obtained from, respectively, {\em friendly} and {\em hostile} merging. We clarify the similarities and differences and suggest that the common overall features can be described in terms of two distinct prototypes of directed scale-free-like networks. Finally we compare with two real directed networks {\em i.e.\ }metabolic networks and transcriptional networks and discuss the similarities and differences.

\section{Friendly and Hostile Merging:}

Recently Kim et al in Ref.\ \cite{Beom05} constructed a local update rule of merging type for {\em undirected} networks which automatically gives rise to scale-free degree distributions. We here extend this local merging evolution to {\em directed} networks and describe and discuss two alternatives called {\em friendly} and {\em hostile} merging.  
\subparagraph{Friendly Merging:}

Friendly merging is an example of a local update which automatically gives rise to scale-freeness and, as we will discuss, a prototype of directed networks. A specific context could, for example, be the situation when companies invest money in other companies. If company A has invested in company B there is a link from A to B. A may for example be a big company with a large turnover represented by a lot of in and out-links, whereas B could be a small company with just a few links. The friendly merging describes how companies buy up each other and new companies are started from scratch until a steady state has been reached where the average size of the companies is constant.

\begin{figure}[!tb]
\includegraphics[width=\columnwidth]{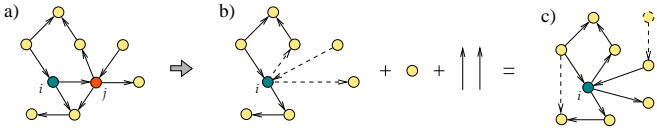}
\centering
\caption{Friendly Merging: a) Node $i$ is randomly picked to merge with one of its out-link neighbors, $j$: b) Node $i$ gets all the links that are connected to $j$ except the ones they have in common and the ones pointing to each other ($N_{common,in}$ and $N_{common,out}$), so one node and two links are taken away from the system shown in the figure. Node $i$ thus gets the in- and out-degree, $\tilde{k}_{i,in}=k_{i,in}+k_{j,in}-N_{common,in}$ and $\tilde{k}_{i,out}=k_{i,out}+k_{j,out}-N_{common,out}$, respectively. c) One node and two links are put in at random to keep the system size constant.}
\label{Fig_UPF}
\end{figure}

The specific local update rule for friendly merging is illustrated in Fig.\ \ref{Fig_UPF} and can be described as follows:
\begin{itemize}
\item[a)] {\em Choose a node}: Randomly pick a node, $i$, with in- and out degree $k_{i,in}$ and $k_{i,out}$.
\item[b)] {\em Choose a node to merge with}: Randomly pick one of its neighbors, $j$, with in- and out-degree $k_{j,in}$ and $k_{j,out}$, through one of the out-links of $i$.
\item[c)] {\em Merging step}: Move all the links (in- and out-links) connected to $j$, so that they connect to $i$ instead. Node $i$ will then have the in- and out-degrees $\tilde{k}_{i,in}= k_{i,in}+ k_{j,in}-N_{common,in}$ and 
$\tilde{k}_{i,out}= k_{i,out}+ k_{j,out}-N_{common,out}$, respectively. Here $N_{common,in/out}$ is the number of links which the nodes $i$ and $j$ have to common neighbors (including the one to each other). Since only at most one link of each direction between two different nodes are allowed, these in- and out-degrees disappear. This reflects the efficiency gain in the merging step.
\item[d)] {\em Balance step}: Add a new node with degree zero,  $k_{l,in}=k_{l,out}=0$, and then add $r$ links that connect randomly into the system. This step ensures that the number of nodes is constant and that the number of links reaches an equilibrium value after going through the updating steps many times. The equilibrium value of the links is thus controlled by the parameter $r$. In equilibrium $r=\langle N_{common,in}+N_{common,out}\rangle$.
\end{itemize}

\begin{figure}[!tb]
\includegraphics[width=\columnwidth]{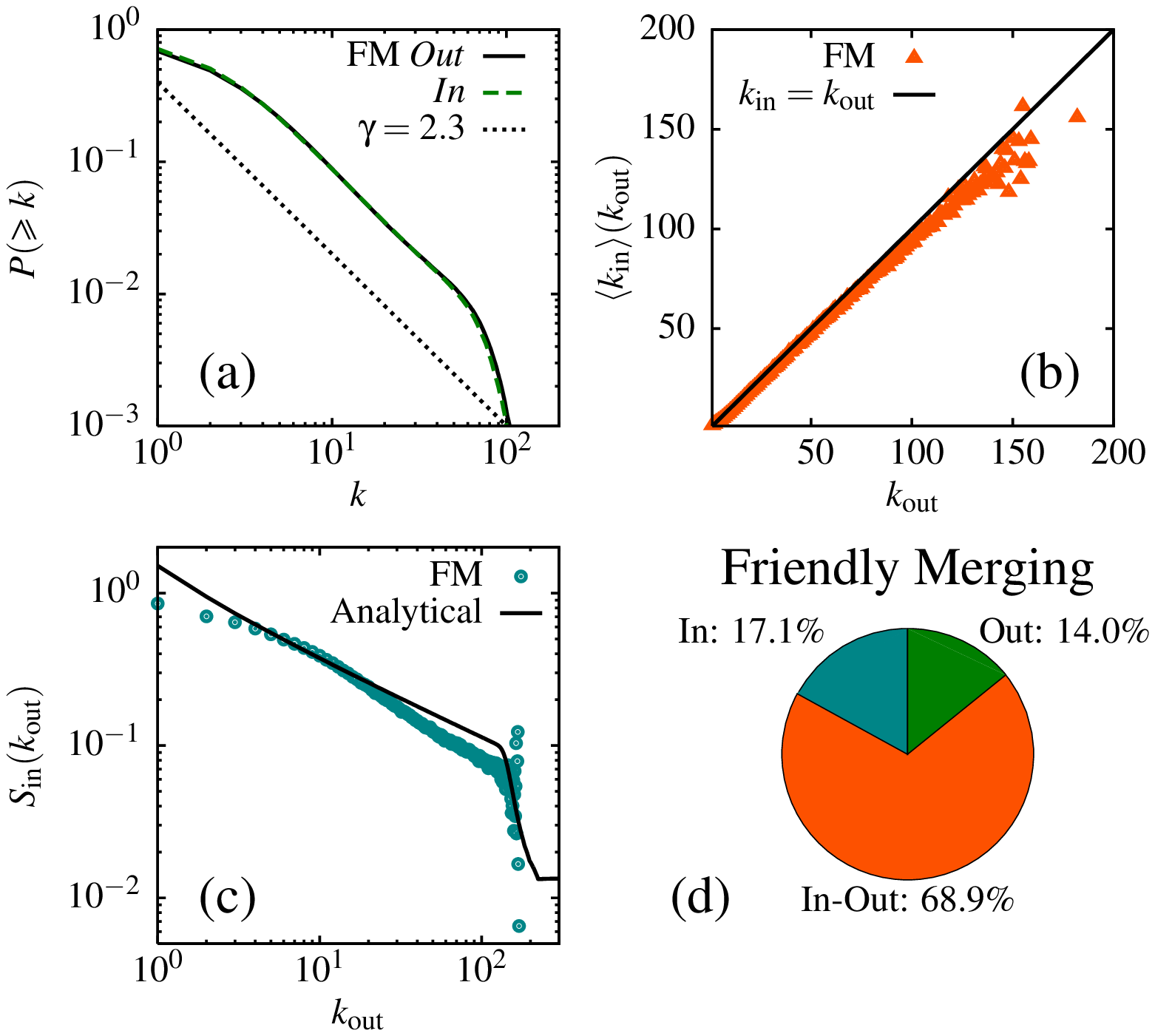}
\centering
\caption{(Color online) Friendly Merging: a) Cumulative degree-distribution $P(\geq k)=\int_k^\infty dkP(k)$ for $\gamma\approx 2.3$ and $\langle k_{in}\rangle =\langle k_{out}\rangle \approx4.0$ (corresponding to $r=3$ in the update rule): b) Demonstration that $\langle k_{in}\rangle_{out}=k_{out}$ to good approximation: c) Demonstration that the spread goes as $S_{in}(k_{out})\propto k_{out}^{-1/2}$ to good approximation; full line Eq.\ (\ref{binom}) and data points from simulations. The cut-off for large $k_{out}$ is a finite size effect. d) Proportion of links with in-, out-, and both in- and out-links.}
\label{Fig_FM}
\end{figure}

Here, $``\langle\rangle"$ denote the average. We start from an Erd\H os-R\'{e}nyi(ER)-network (see e.g.\ \cite{Dorog02}) and apply the merging update until equilibrium is obtained. We use the size $N = 1000$-nodes for the simulations of model networks in order to facilitate comparison with the real networks in the paper, which have approximately this size. For $r=3$ we obtain $\langle k_{in}\rangle=\langle k_{out}\rangle\approx4$.
Figure \ref{Fig_FM} gives the characteristics of this merging network obtained by averaging over many equilibrium networks (for the merging models we use 100 networks whereas in model A and B we use 10000 networks).
As demonstrated by Fig.\ \ref{Fig_FM}(a), the in- and out-degree distributions $P_{out}(k_{out}=P_{in}(k_{in})$ are equal and ``scale-free" like, {\em i.e.\ }can be approximated by a power law over a substantial range (see Fig.\ \ref{Fig_FM}(a)) \cite{Jeong00}. The scale-freeness is a truly non-trivial emergent property whereas the fact that $P_{out}(k_{out}$ is equal to $P_{in}(k_{in})$ basically reflects the symmetry between in- and out-links in the friendly merging  scheme (note however the slight asymmetry introduced by choosing the neighbor link from an out-link, discussed in connection with Fig.\ \ref{Fig_FM}(d)). Fig.\ \ref{Fig_FM}(b) demonstrates a second non-trivial property: the average number of in-degrees $\langle k_{in}\rangle$ for nodes with a fixed out-degree $k_{out}$ is closely equal to $k_{out}$. By contrast the random ER-network gives a horizontal line $\langle k_{in}\rangle_{out}=\langle k_{in}\rangle$.

In order to get some insight into the origin of this second non-trivial property we introduce the concept of a $random$ scale-free network: The assumption is that the total number of links $k_{in}+k_{out}$ on a node is given by the same distribution as $k_{in}$ and $k_{out}$. This assumption is fulfilled in case of friendly merging. Next we assume that the $numbers$ of in- and out links on a $node$ of size $k$ are randomly distributed and that the distribution for $k$ is scale-free. The relation between in-and out-degrees on a node, can then be understood by ignoring the constraint implicit in having a connected network. We instead consider the nodes as boxes of certain sizes (degrees), $k$, and with the size distribution $P(k)$. We then put in red and blue balls (respectively, in- and out-links) into the boxes until they are filled. In such a case the average number of in-links (red balls) on a node with precisely $k_{\rm out}$ out-links (blue balls), $\langle k_{in}\rangle_{k_{\rm out}}$, is just given in terms of the binomial distribution $B(k_{in},k)$ i.e.\ the probability to get $k_{in}$ tails when tossing a coin $k=k_{\rm in}+k_{\rm out}$ times:

\begin{equation}
\label{kin}
\langle k_{in}\rangle_{k_{\rm out}}=\frac{\sum_{k=k_{out}}^{k_{max}} B(k-k_{out},k)P(k)(k-k_{out})}{\sum_{k=k_{ out}}^{k_{max}}B(k-k_{out},k)P(k)}
\end{equation}
where $k$ is the total number of links attached to a node (so that $k_{\rm in}=k-k_{out}$), and $k_{max}$ is the largest node in the network. Here $P(k=k_{\rm in}+k_{\rm out})$ is the probability of picking a node of size $k$. For a the case of $P(k)\propto 1/k^\gamma$ one then finds $\langle k_{\rm in}\rangle \approx k_{\rm out}$ (the analytical solution for $\gamma=2$ gives $\langle k_{\rm in}\rangle=k_{\rm out}-2$ to leading order of large $k_{\rm out}$).

We pursue this property of random scale-free networks one step further and study the spread of the $k_{in}$-links for the nodes with a given number of $k_{out}$. For this spread we use the measure
\begin{equation}
S_{in}(k_{out})=\frac{\sum_{(k_{in}|k_{out})}|k_{in}-\langle k_{in}\rangle_{k_{out}}|}{N_{k_{out}}\langle k_{ in}\rangle _{k_{out}}}
\label{Sin}
\end{equation}
Using the same simplification as in Eq.\ (\ref{kin}) then gives 

\begin{eqnarray}
&&S_{in}(k_{out})=\nonumber\\
&&\frac{\sum_{k=k_{out}}^{k_{max}}B(k-k_{out},k)P(k)\frac{|k-k_{out}-\langle k_{in}\rangle_{k_{out}}|}{\langle k_{in}\rangle_{k_{out}}}}{\sum_{k=k_{out}}^{k_{max}}B(k-k_{\rm out},k)P(k)}
\label{binom}
\end{eqnarray}

\begin{figure}[!tb]
\includegraphics[width=\columnwidth]{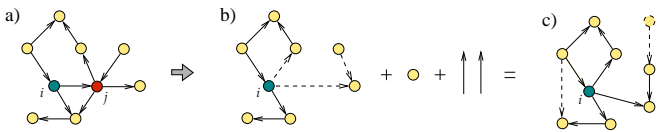}
\centering
\caption{(Color online) Hostile Merging: a) Node $i$ is randomly picked to merge with one of its out-links neighbors, $j$. b) Node $i$ gets all the out -links that are connected to $j$ except the ones $i$ and $j$  have in common neighbors. The in-links to $j$ are randomly rewired to other nodes (excluding $i$). This means that one node and two links are taken away from the system. Node $i$ thus gets the in- and out-degree, $\tilde{k}_{i,in}=k_{i,in}$ and $\tilde{k}_{i,out}=k_{i,out}+k_{j,out}-N_{common,out}$. c) One node and two links are put in at random to keep the system size constant.}
\label{Host_merg}
\end{figure}

For a random scale-free distribution this gives $S_{\rm in}(k_{\rm out})\propto k^{-1/2}_{\rm out}$,whereas for the ER-network the spread is independent of $k_{\rm out}$ i.e. $S_{\rm in}(k_{\rm out})=const$. The spread obtained from friendly merging is plotted in Fig.\ \ref{Fig_FM}(c) (with the use of Eq.\ \ref{Sin}) together with the analytical result obtained from Eq.\ (\ref{binom}). The sharp change in the slope of the analytical result in Fig.\ \ref{Fig_FM}(c), at large $k_{out}$, is a finite size effect. The drop occures when $k_{out} \approx k_{max}/2$.
Figures \ref{Fig_FM} (a)-(c) show that friendly merging essentially gives rise to a {\em random} scale-free network.

Finally in Fig.\ \ref{Fig_FM}d we give the percentage of nodes with only in-links, nodes with only out-links and nodes with both in- and out-links. One notes that friendly merging gives rise to a slightly larger percentage of nodes with only in-links than with only out-links. This asymmetry is caused by always picking an out-link neighbor in the update rule a) for Friendly merging. Among the nodes with both in- and out-links a total of 12 \% have double links (a double link e.g. means that company A and B has mutually invested in each other).

This constitutes our description of friendly merging and the characteristics of the directed network it gives rise to. Next we turn to an alternative local update rule termed hostile merging.

\subparagraph{Hostile Merging:}
The alternative update rule, hostile merging, can also be described in terms of the company analogy: In this case company A makes a hostile takeover and acquires all the assets of company B. The companies that had money invested in B prior to the takeover will not be allowed to have any control over A. So in this case these companies will be forced to sell their parts in company B to A and invest the money elsewhere. In terms of networks this means that a node gets all the out-links from the neighbor it merges with. This can be translated into the following update rule:

\begin{figure}[!tb]
\includegraphics[width=\columnwidth]{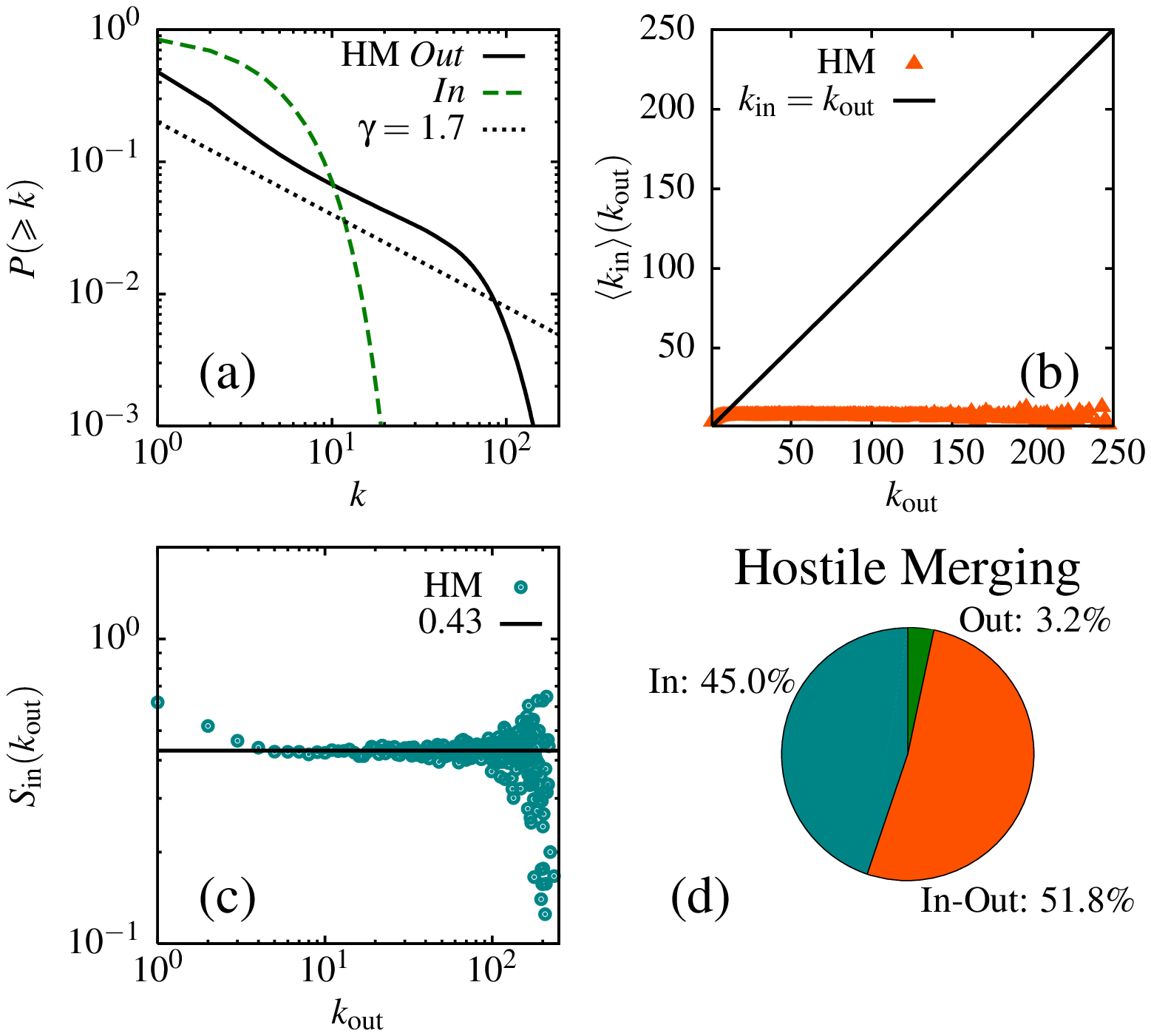}
\centering
\caption{(Color online) Hostile Merging: a) Cumulative degree-distribution $P_{out}(\geq k_{out})$ and $P_{in}(\geq k_{in})$ obtained for $\langle k\rangle \approx 3.8$ (this particular value was chosen in order to facilitate comparison with the yeast network in Fig.\ \ref{Fig_YPD}) which corresponds to $r=1.5$ in the local update (the dotted straight line has the slope $-1.7$). b) No correlations between $\langle k_{in}\rangle_{out}$ and $k_{out}$ for the $k_{out}$-nodes because $\langle k_{in}\rangle_{out}=\langle k_{in}\rangle$ independent of $k_{out}$. c) Also the spread $S_{in}(k_{out})$ is independent of $k_{out}$. d) Proportion of links with in-, out-, and both in- and out-links.}
\label{Fig_HM}
\end{figure}

\begin{itemize}
\item[a)] {\em Choose a node}: Same as for friendly merging.
\item[b)] {\em Choose a node to merge with}: Same as for friendly merging.
\item[c)] {\em Merging step}: Move all the out-links connected to $j$, so that they connect to $i$ instead and move all the in-links of $j$ so that they randomly connect to other nodes (excluding $i$). Links sitting between $i$ and $j$ before the merging are deleted. Node $i$ will then have the in- and out degrees $\tilde{k}_{i,in}= k_{i,in}$ (or $\tilde{k}_{i,in}= k_{i,in}-1$, if a link from $j$ to $i$ existed) and
$\tilde{k}_{i,out}= k_{i,out}+ k_{j,out}-N_{common,out}$, respectively. Here $N_{common,out}$ is the number of out-links which the nodes $i$ and $j$ have to the same neighbors.
\item[d)] {\em Balance step}: Same as for friendly merging.
\end{itemize}

The hostile merging update rule is illustrated in Fig.\ \ref{Host_merg} and in Fig.\ \ref{Fig_HM} we present the characteristic features for the network emerging from this evolution rule.

As seen from Fig.\ \ref{Fig_HM} networks arising from hostile merging are characterized by a broad "scale-free"-like out-degree distribution and an ER-like in-degree distribution with no correlation between the number of in- and out-degrees on the nodes with a given number of in- or out-degrees. There is again  a large portion of links with both out-and in-degrees (52\% as compared to 69\% for friendly merging). However the asymmetry between links with only in- and only out-degrees is huge for hostile merging (45\% for only in and 3.2\% for only out) and the proportion of nodes with double links is 14\%.

The networks arising from hostile merging and friendly merging are thus very different. The common feature is basically that the out-degree distribution is scale-free-like in both cases. We will in the following argue that the two different types of networks arising from friendly and hostile merging can in fact be viewed as two prototypes of directed networks connected to scale-freeness.

\section{Model A and B}
So far we have constructed two local evolution rules and found that the two types of emerging directed networks both display scale-free features but are otherwise different. In this section we show that the overall features of these two network types can be connected to two minimalistic random scale-free network-models. We suggest that these two minimalistic network models can be viewed as two prototype models. 

\subparagraph{Model A:}
\begin{figure}[!tb]
\includegraphics[width=\columnwidth]{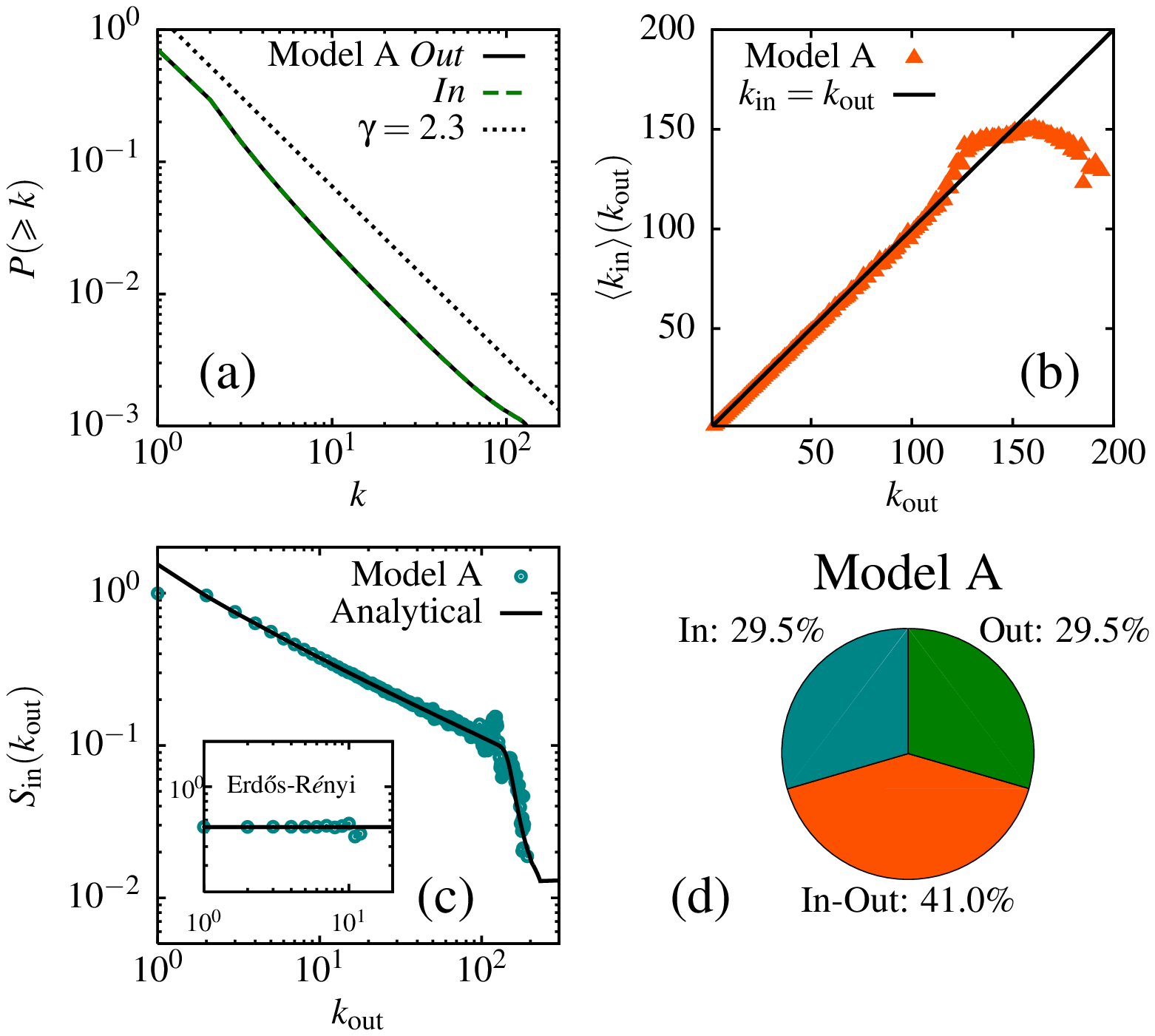}
\centering
\caption{(Color online) Model A: a) Cumulative degree-distribution $P(\geq k)=\int_k^\infty dkP(k)$ for $\gamma\approx 2.3$ and $\langle k\rangle \approx1.9$: b) Demonstration that $\langle k_{in}\rangle_{out}=k_{out}$. (An ER-network has instead $\langle k_{in}\rangle_{out}=\langle k\rangle /2$). c) Demonstration that the spread goes as $S_{in}(k_{out})\propto k_{out}^{-1/2}$. full line Eq.\ (\ref{binom}) and data points from simulations. The cut-off for large $k_{out}$ is a finite size effect. d) Proportion of links with in-, out-, and both in- and out-links.}
\label{Fig_StdA}
\end{figure}

Model A is constructed as follows: We start from a scale-free undirected connected network with a degree-distribution $P(k)\propto k^{-\gamma}$ and with average degree $\langle k\rangle$ (which means that in the limit of very large systems $\langle k\rangle=(\gamma -1)/(\gamma -2)$). This scale-free undirected network is constructed by the Stub algorithm followed by a random rewiring \cite{Newman01,Maslov02,Maslov04}. Next we randomly assign directions to links with equal probability. Such a network has the following obvious property: The in-degree and the out-degree distributions are equal, $P(k_{in})= P(k_{out})$ and both proportional to $k^{-\gamma}$. Finally, we separately rewire the in- and out-end of the links randomly without changing the scale-free distributions. Figure \ref{Fig_StdA}(a) shows that the resulting distributions are indeed equal and scale free. Figure \ref{Fig_StdA}(b) demonstrates that $\langle k_{\rm in}\rangle \approx k_{\rm out}$ (the deviation for large  $k_{\rm out}$ is a finite size effect). This explicitly links this relation to networks where the in- and out-degree distributions are both scale-free and equal and an average is taken over a large ensamble of networks (as explained in connection with Eq.\ (\ref{kin})). In the same way Fig.\ \ref{Fig_StdA}(c) connects the spread predicted by Eq.\ (\ref{binom}) for equal scale-free in- and out-degree distributions  to the result obtained from simulation of model A (full curve in Fig.\ \ref{Fig_StdA}(c), the inset shows the corresponding result for an ER-network). We find that model A is a random minimalistic model with the essential characteristics of the friendly merging networks (compare Fig.\ \ref{Fig_FM}). Figure \ref{Fig_StdA}(d) gives the proportions of nodes with only in-, only out- and both in- and out-links. Comparing with friendly merging in Fig.\ \ref{Fig_FM}(d), we notice that the number of only in- and only out-nodes is somewhat larger for Model A and that there is a perfect symmetry in contrast to the slight asymmetry discussed in connection with friendly merging. Our conclusion is hence that friendly merging shares its {\em overall} characteristics with model A.

\subparagraph{Model B:}

\begin{figure}[!tb]
\includegraphics[width=\columnwidth]{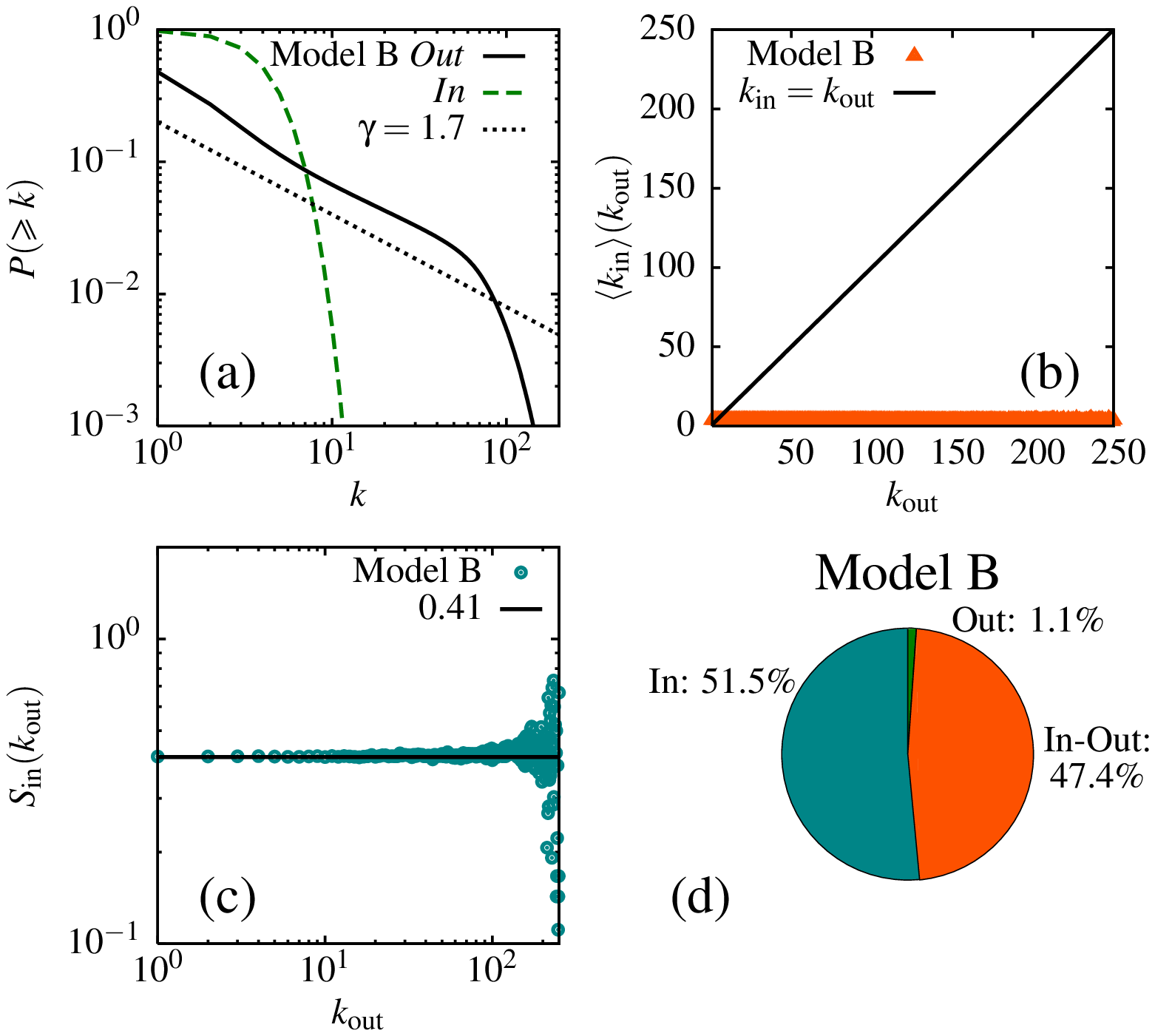}
\centering
\caption{(Color online) Model B: a) Cumulative scale-free degree-distribution $P_{out}(\geq k_{out})$ for $\gamma\approx 1.7$ and $\langle k_{out}\rangle \approx3.8$ together with the ER-like $P_{in}(\geq k_{in})$. b) No correlations because $\langle k_{in}\rangle_{out}=\langle k_{in}\rangle$ independent of $k_{out}$. c) The spread $S_{in}(k_{out})$ independent of $k_{out}$ for the same reason. d) Proportion of links with in-, out-, and both in- and out-links.}
\label{Fig_StdB}
\end{figure}

The minimalistic model B, displaying common characteristics with the hostile merging networks is constructed as follows: We again start from a scale free distribution of the total node degree $P(k)\propto k^{-\gamma}$ (the total degree distribution is the same as for hostile merging for better comparison) and randomly assign directions on the links. Next we make a random rewiring on the links keeping the out-degree distribution intact, but rewiring without this restriction for the in-degrees. This means that $P(k_{out})\propto  k_{out}^{-\gamma}$ whereas the in-degree distribution becomes ER-like. Model B is consequently by construction a prototype of a random model where the out-degree is scale-free and the in-degree is of ER-type.

The general characteristics of model B is shown in Fig.\ \ref{Fig_StdB} for $\gamma=1.7$ and $\langle k\rangle \approx3.8$. Figure \ref{Fig_StdB}(a) shows the scale-free broad out-degree distribution and the narrow ER-type in-degree distribution. Figure \ref{Fig_StdB}(b) illustrates that there is no correlation between in and out-degrees for nodes with a given fixed number of $k_{out}$. This follows because by construction the average in-degree on any node is $\langle k_{in}\rangle=\langle k\rangle/2$ regardless of the number of out-links (compare Model A and Fig.\ \ref{Fig_StdA}(b)). As a consequence the spread in Fig.\ \ref{Fig_StdB}(c) is independent of $k_{out}$ (compare Model A and Fig.\ \ref{Fig_StdA}(c)). Finally, Fig.\ \ref{Fig_StdB}(d) shows the percentage of in- and out-degrees on the nodes. 
By comparing Fig.\ \ref{Fig_StdB} with Fig.\ \ref{Fig_HM}, we find that model B catches the {\em overall} features of the hostile merging networks. 

\section{Metabolic and Transcriptional networks}
Do the two prototypical properties A and B found in respectively friendly and hostile merging also show up in real networks? We here demonstrate that they indeed do, with A showing up in biological production networks, i.e. metabolic networks, while B rather is found in information processing networks like the transcriptional networks.

\subparagraph{Metabolic networks:}

\begin{figure}[!tb]
\includegraphics[width=\columnwidth]{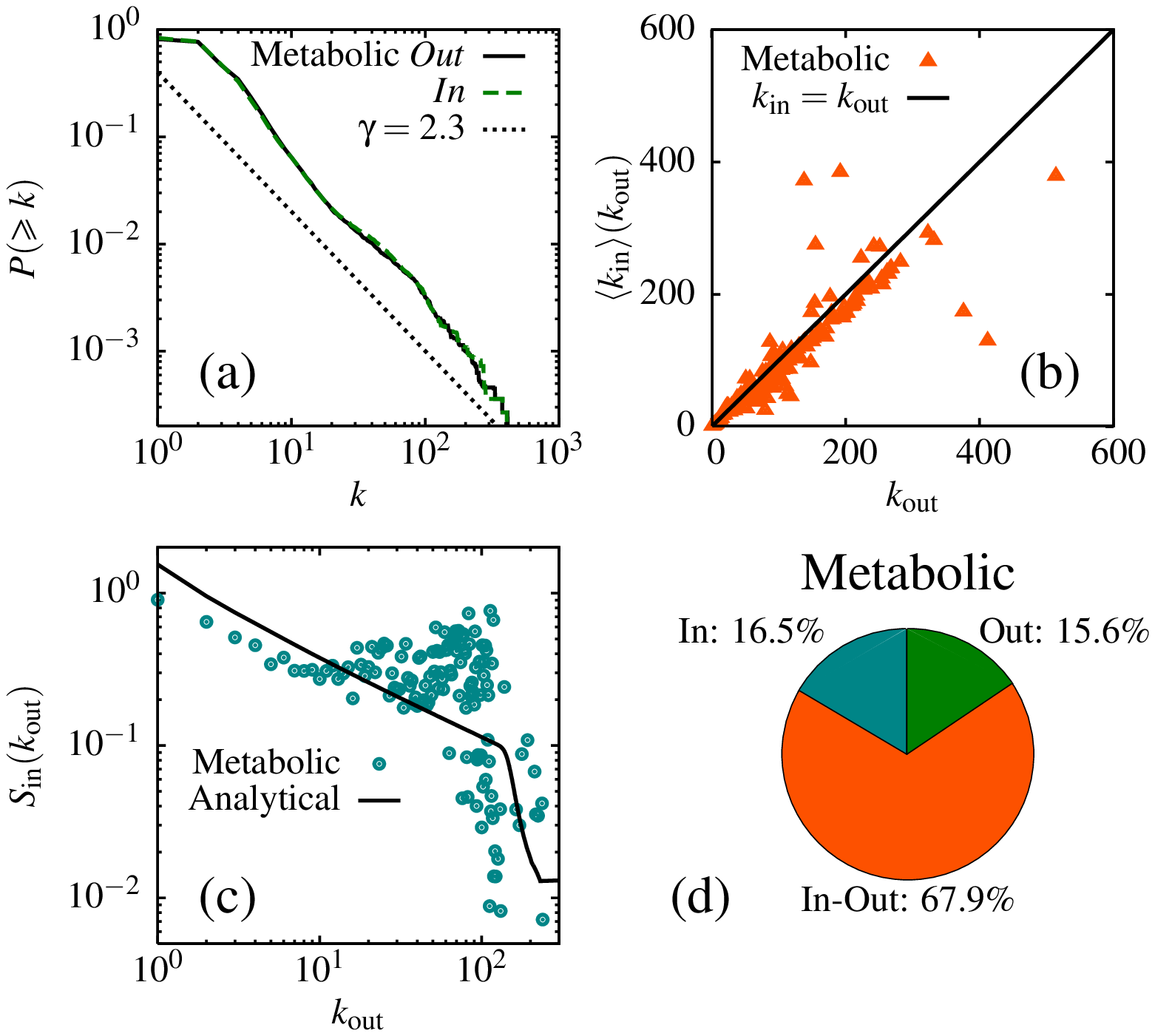}
\centering
\caption{(Color online) Metabolic Networks: Average over 107 metabolic networks with data obtained from Ref.\ (\cite{Ma03a}). The same characteristics as for Model A and  Friendly Merging: a) Cumulative degree-distribution $P(\geq k)$. The dashed straight line line has the slope $\gamma=2.3$ and $\langle k_{in}\rangle=\langle k_{out}\rangle \approx4.3$. b) Plot of $\langle k_{in}\rangle_{out}$ versus $k_{out}$ showing that the data is consistent with $\langle k_{in}\rangle_{out}=k_{out}$. c) Demonstration that the spread goes as $S_{in}(k_{out})\propto k_{out}^{-1/2}$ to reasonable approximation, full line from Eq.\ (\ref{binom}) and data points from simulations. The cut-off for large $k_{out}$ is a finite size effect. d) Proportion of links with in-, out-, and both in- and out-links.}
\label{Fig_Metab}
\end{figure}

The first example is the average properties of 107 metabolic networks with the average size $\langle N\rangle \approx940$ (data taken from Ref.\ \cite{Ma03a}). A metabolic network is constructed as follows: Substrates and products in the metabolism are nodes. Two nodes are connected if the first substance is a substrate in a metabolic reaction which produces the other substance. The links points from the substrate to the product.
The data are obtained as the average over 107 such networks and consequently reflect an ensemble average network structure associated with metabolic networks \cite{expl}. Figure \ref{Fig_Metab}(a) shows that, just as for friendly merging and model A, metabolic networks have $P_{out}(k_{out})=P_{in}(k_{in})$ and has a broad scale-free degree-distribution as was first demonstrated in Ref.\ \cite{Jeong00}. Furthermore from Fig.\ \ref{Fig_Metab}(b) and (c) we conclude that also for the ensemble average of metabolic networks to good approximation the relation $\langle k_{in}\rangle \approx k_{out}$ holds. Also the spread has a similar decrease as in case of friendly merging and model A. In addition the relative proportions of links on a node corresponds very well for metabolic networks and friendly merging (compare Fig.\ref{Fig_Metab}(d) and Fig.\ref{Fig_FM}(d) and notice that these two networks also have approximately the same number of average links). We conclude that the {\em overall} structure of metabolic networks belong to the same network class as A and friendly merging.

\begin{figure}[!tb]
\includegraphics[width=\columnwidth]{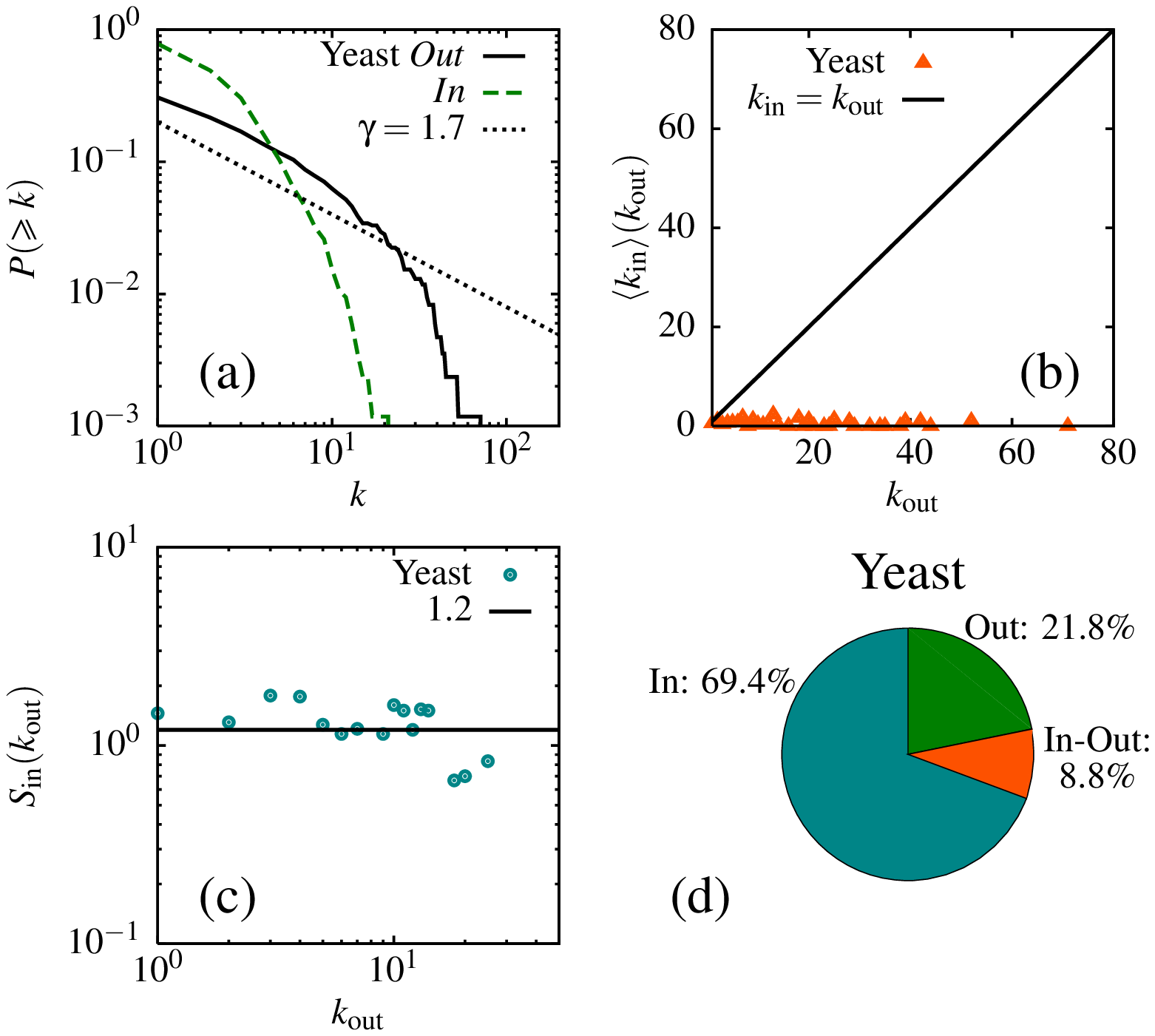}
\centering
\caption{(Color online) Transcription networks for yeast (data from Ref.\ \cite{Hodges99,Hodges98}). The same characteristics as for Model B and  Hostile Merging. a) Cumulative degree-distribution $P_{out}(\geq k_{out})$ and $P_{in}(\geq k_{in})$ showing a broad distribution for the out-degrees (the dotted line is for comparison with hostile merging and has the slope $\gamma\approx 1.7$) and an ER-like distribution of in-links. b) Shows that there are no correlations between $\langle k_{in}\rangle_{out}$ and $k_{out}$ for the $k_{out}$-nodes. c) Also the spread $S_{in}(k_{out})$ is independent of $k_{out}$. d) Proportion of links with in-, out-, and both in- and out-links.}
\label{Fig_YPD}
\end{figure}

\subparagraph{Transcriptional networks:}
Figure \ref{Fig_YPD} shows the corresponding analysis for the network of transcriptional protein-protein regulations for  yeast(Saccromyces Cerevisiae, data from Ref.\ \cite{Hodges99,Hodges98}). Comparing Figs.\ \ref{Fig_YPD}(a)-(c), \ref{Fig_HM}(a)-(c) and \ref{Fig_StdA}(a)-(c), shows the common feature of a broad out-degree and an ER-like in-degree. The ER-like character for the in-degree of yeast is emphasized by the lack of correlations displayed by Figs.\ \ref{Fig_YPD}(b) and (c). Overall we see that transcriptional networks belongs to the class characterized by model B. However, there are also differences: the out-degree for the yeast-network is broad but not very scale-free and the percentages of degrees in Fig.\ \ref{Fig_HM}(d) on the one hand are significantly different as compared to hostile merging and model B on the other hand.

\section{Conclusions}
We present and characterize two directed network-types emerging from two different local time-evolutions called friendly and hostile merging. It was shown that two minimalistic models contain the same overall characteristics as, respectively, the friendly and hostile merging networks. We compared metabolic and transcription networks from real data with the two prototypes and found that the properties of metabolic networks has the same overall characteristics as model A and the friendly merging network. However, the yeast transcriptional network resembled model B and hostile merging networks. This led us to suggest that friendly- and hostile merging networks represent two distinct classes of directed networks.

What might be the implication of these results? Even if friendly merging and metabolic networks are similar, we do not  suggest that the friendly merging evolution rule (which may be motivated in an economical context) has direct correspondence to the actual evolution of metabolic networks (which has to do with production handling). The implication is rather that scale freeness can arise in a variety of explicit ways. However, it was also shown that an ensemble of {\em random} scale-free networks bring additional non-trivial properties like a relation between in- and out-links on a node. Comparison between the ensemble average of metabolic networks implies that this latter property is in fact shared with the metabolic networks. This imposes restrictions on the evolution and the correlations of individual metabolic networks.

Since scale-freeness is a common property and since ensemble averages of real directed networks show strong similarities with random scale-free networks, there obviously is a need for further distinguishing network measures. In the present paper we suggest and discuss one such measure: The percentage of nodes of different number of only in-, only out-, both in- and out-links.\cite{Gronlund04} We found a close correspondence between the friendly merging network and the ensemble average of metabolic networks.

In summary we uncovered two distinct classes of directed networks A and B. In case of friendly and hostile merging we suggest that the distinction between A and B might reflect a difference in economical strategy. For biological networks we suggest that the distinction instead reflects a functional design difference associated with handling production (metabolic networks) and information processing (transcriptional networks). 

\subsection*{Acknowledgment}
Kim Sneppen and Martin Rosvall are thanked for stimulating discussions and comments. Support from the Swedish Research Council grant 621-2002-4135 is greatfully acknowledged.


\begin{thebibliography}{19}
\providecommand{\natexlab}[1]{#1}
\providecommand{\url}[1]{\texttt{#1}}
\expandafter\ifx\csname urlstyle\endcsname\relax
  \providecommand{\doi}[1]{doi: #1}\else
  \providecommand{\doi}{doi: \begingroup \urlstyle{rm}\Url}\fi

\bibitem[Albert and Barab\'{a}si(2002)]{Albert02}
R.~Albert and A.-L Barab\'{a}si.
\newblock \emph{Rev.Mod. Phys.}, 74:\penalty0 47, 2002.

\bibitem[Boccaletti et~al.(2006)Boccaletti, Latora, Morena, Chavez, and
  Hwang]{Bocca06}
S.~Boccaletti, V.~Latora, Y.~Morena, M.~Chavez, and D.-U. Hwang.
\newblock \emph{Phys. Rep.}, 424:\penalty0 175, 2006.

\bibitem[Dorogovtsev and Mendes(2002)]{Dorog02}
S.N. Dorogovtsev and J.F.F. Mendes.
\newblock \emph{Adv.Phys.}, 51:\penalty0 1079, 2002.

\bibitem[Newman(2003)]{Newman03}
M.E.J. Newman.
\newblock \emph{SIAM Review}, 45:\penalty0 167, 2003.

\bibitem[Strogatz(2001)]{Strogatz01}
S.H. Strogatz.
\newblock \emph{Nature}, 410:\penalty0 268, 2001.

\bibitem[Barab\'{a}si(2000)]{Barhome}
A.-L. Barab\'{a}si.
\newblock http://www.nd.edu/networks/resources.htlm.
\newblock 2000.

\bibitem[Jeong et~al.(2000)Jeong, Tombor, Albert, Oltvai, and
  Barab\'{a}si]{Jeong00}
H.~Jeong, B.~Tombor, R.~Albert, Z.N. Oltvai, and A.-L Barab\'{a}si.
\newblock \emph{Nature}, 407:\penalty0 651, 2000.

\bibitem[Ma and Zeng(2003{\natexlab{a}})]{Ma03a}
H.~Ma and A.-P Zeng.
\newblock \emph{Bioinformatics}, 19:\penalty0 270, 2003{\natexlab{a}}.

\bibitem[Ma and Zeng(2003{\natexlab{b}})]{Ma03b}
H.~Ma and A.-P Zeng.
\newblock \emph{Bioinformatics}, 19:\penalty0 1423, 2003{\natexlab{b}}.

\bibitem[Tanaka(2005)]{Tanaka05}
R.~Tanaka.
\newblock \emph{Phys. Rev. Lett.}, 94:\penalty0 1681011, 2005.

\bibitem[Hodges et~al.(1998)Hodges, Payne, and Garrels]{Hodges99}
P.E. Hodges, W.E. Payne, and J.I. Garrels.
\newblock \emph{Nucl. Acids Res}, 26:\penalty0 68, 1998.

\bibitem[Hodges et~al.(1999)Hodges, McKee, Davis, Payne, and Garrels]{Hodges98}
P.E. Hodges, A.H. McKee, B.P. Davis, W.E. Payne, and J.I. Garrels.
\newblock \emph{Nucl.Acids Res}, 27:\penalty0 69, 1999.

\bibitem[Yook et~al.(2004)Yook, Oltvai, and Barab\'{a}si]{Yook04}
S.-H. Yook, Z.N. Oltvai, and A.-L Barab\'{a}si.
\newblock \emph{Proteomics}, 4:\penalty0 928, 2004.

\bibitem[Kim et~al.(2004)Kim, A.~Trusina, and Sneppen]{Beom05}
B.J. Kim, P.~Minnhagen A.~Trusina, and K.~Sneppen.
\newblock \emph{Eur.Phys.J.B}, 43:\penalty0 369, 2004.

\bibitem[Maslov and Sneppen(2002)]{Maslov02}
S.~Maslov and K.~Sneppen.
\newblock \emph{Science}, 296:\penalty0 910, 2002.

\bibitem[Maslov et~al.(2004)Maslov, Sneppen, and Zaliznyak]{Maslov04}
S.~Maslov, K.~Sneppen, and A.~Zaliznyak.
\newblock \emph{Physica A}, 333:\penalty0 529, 2004.

\bibitem[M.E.J.~Newman(2001)]{Newman01}
D.J.~Watts M.E.J.~Newman, S.H.~Strogats.
\newblock \emph{Phys.Rev.E}, 64:\penalty0 026118, 2001.

\bibitem[Ex(2006)]{expl}
Ensemble average here refers to that each quantity is averaged over all the nodes in the total ensemble. The assumption made is that the differences between various metabolic networks for the quantity analysed are to large extent statistical in nature. A deeper analysis to verify this assumption will require a larger data set of metabolic networks.

\bibitem[Gr\"{o}nlund(2003)]{Gronlund04}
A.~Gr\"{o}nlund.
\newblock \emph{Phys. Rev E}, 70:\penalty0 061908, 2003.

\end{thebibliography}
\end{document}